\documentclass{mn2e}
\input{epsf}

\title[Angular distribution of satellite galaxies from SDSS DR4]
{Angular~distribution~of~satellite~galaxies from the Sloan~Digital~
Sky~Survey~Data~Release~4}
\author[Azzaro et al.]{Marco Azzaro$^{1}$, Santiago G. Patiri$^{2}$, Francisco Prada$^{3}$, 
\& Andrew R. Zentner$^{4}$\\ 
\\
$^1$ Universidad de Granada,
Av. Fuentenueva s/n, E-18071 Granada, Spain ({\rm azzaro@ugr.es})\\
$^2$ Instituto de Astrof{\'\i}sica de Canarias,
C/V{\'\i}a L\'actea s/n, E-38200, La Laguna, Tenerife, Spain\\
$^3$ Instituto de Astrof{\'\i}sica de Andalucia (CSIC),
Apartado Correos 3005, E-18080 Granada, Spain\\
$^4$ Kavli Institute for Cosmological Physics \& Department of Astronomy and Astrophysics,\\
\hspace*{0.1cm} The University of Chicago, 933 E. 56 St., Chicago, IL 60637, USA}

\begin{document}

\maketitle

\vspace{-1in}

\begin{abstract}
We explore the angular distribution of two samples of satellite galaxies
orbiting isolated hosts extracted from the Sloan Digital Sky Survey Data
Release 4. We find a clear alignment of the satellites along the major
axis of their hosts when restricting the analysis to red hosts.  The 
anisotropy is most pronounced for red satellites of red hosts.  
We find that the distribution of the satellites about blue, isolated hosts 
is consistent with isotropy.  We show that under the {\it assumption} that 
the true, underlying distribution of satellites of blue 
hosts exhibits the same anisotropy as the satellites of red hosts, 
the sample of blue hosts is too small to 
measure this anisotropy at a statistically-significant level.  
The anisotropy that we detect for satellites about red primaries is 
independent of the projected radius.  In particular, it is evident at 
large projected distances from primaries ($300 < r_{p} < 500$\,kpc).

\end{abstract}

\begin{keywords}
galaxies: general -- galaxies: formation -- galaxies: evolution\\
--  galaxies: interactions -- galaxies: kinematics and dynamics
\end{keywords}

\vspace{-0.45in}

\section{Introduction}

In the standard hierarchical cold dark matter (CDM) model 
of cosmological structure and galaxy formation 
(e.g., White \& Rees, 1978, Blumenthal et al. 1984),
the halos of large galaxies form by the accretion 
of smaller halos.  During this process, 
some accreted halos may orbit
several times within the {\em host} potential before 
being incorporated into the central galaxy 
or severely disturbed by tides
(Taffoni et al. 2003, Hayashi et al. 2003, Zentner and
Bullock 2003, Kravtsov et al. 2004, Kazantzidis et al. 2004,
Taylor Babul 2004, Zentner et al. 2005).  
The angular distribution of satellites around central hosts
carries the imprint of both anisotropic infall 
(Knebe et al. 2004; Zentner et al. 2005b) 
during halo formation and the shape of the 
gravitational potential on the scale of 
central halos (e.g., Zentner et al. 2005b).  

The angular distribution of satellite galaxies has received 
much recent attention.  
The topic was first addressed by Holmberg (1969), 
who found that satellites of spiral galaxies with 
projected separations $r_{p} \leq 50$\,kpc are 
preferentially located near the short axes of 
their host galaxies.  Zaritsky et al. (1997) found a 
statistically-significant anisotropy in the same 
sense as that of Holmberg (1969), but at larger projected 
separations ($200$\,kpc $\leq r_{p} \leq 500$\,kpc).  
The satellite galaxies of the Local Group exhibit a similar anisotropy 
(e.g., Lynden-Bell 1982; Majewski 1994; Hartwick 1996), 
and this may have important consequences for galaxy formation theory 
(e.g., Kang et al. 2005; Zentner et al. 2005b; Libeskind et al. 2005).

Sales \& Lambas (2004) studied the alignment of satellites with 
respect to the principal axes of the light distributions of their 
host galaxies in the Two Degree Field Galaxy Redshift Survey 
(2dFGRS, Colless et al. 2001).  Sales \& Lambas (2004) reported a 
statistically-significant alignment along the minor axes of host galaxies; 
however, Yang et al. (2006) have since shown that this result was due
to a misinterpretation of the Position Angle definition in
the 2dFGRS database and that the alignment is actually along 
host {\it major} axes.  
Moreover, the Sales \& Lambas (2004) alignment was
only in a subsample of hosts with little active star formation 
and satellites with small velocities relative to their hosts, 
$\Delta V < 160 $\,km\,s$^{-1}$.  For a subsample with 
larger relative velocities ($\Delta V > 160$\,km\,s$^{-1}$), 
the results of Sales \& Lambas (2004) are consistent with isotropy.

Brainerd (2005) studied satellites in the Sloan Digital Sky Survey Data
Release 3 (SDSS DR3, e.g. York et al. 2000; Strauss et al. 2002) and also
found a preferential alignment of satellites with the 
major axes of the light distributions of their hosts for $r_{p} < 100$\,kpc.
Brainerd (2005) found their signal to decrease with $r_{p}$, 
reporting a distribution consistent with isotropy for $r_{p} > 250$\,kpc.
Brainerd (2005) did not study any dependence upon the 
spectral or morphological types of the host galaxies.
Yang et al. (2006), using groups of galaxies from the SDSS DR2 
identified via a halo-based group finder from which 
halo properties for groups are inferred 
(Weinmann et al. 2005; Yang et al. 2005), found an 
alignment of satellites with host major axes, becoming 
more pronounced at small separations.  
The strongest alignment was exhibited by red satellites of 
red hosts and by satellites in massive halos.  
Azzaro et al. (2005) studied the distribution of satellites  
in a small subsample of the SDSS sample of Prada et al. (2003) limited 
to hosts that were morphologically identified as disk galaxies and 
found no evidence for anisotropy, though the requirement of a 
morphologically-identified disk primary drove their sample size to 
be significantly smaller than those of Sales \& Lambas (2004), 
Brainerd (2005), and Yang et al. (2006).

Our aim is to study the angular distribution of satellites
about isolated host galaxies with the increased statistics 
afforded by the SDSS fourth data release (DR4).  Our study complements 
that of Yang et al. (2006), in the sense that we focus on 
isolated hosts instead of galaxy groups.
%
%
We refer to central host galaxies as ``primary galaxies'' 
or simply ``primaries.''  Provided primaries are selected in a 
consistent and homogeneous way, it is useful to consider all 
satellites as satellites of a single, fictitious primary in order to 
overcome the small numbers of satellites about individual primaries.  

\vspace{-0.15in}

\section{Data and selection criteria}
%
The data we use are from the latest SDSS public release, DR4.
The SDSS is the largest photometric and spectroscopic survey available.  
CCD imaging covers five colors and spectroscopic follow-up is 
performed down to $M_{r} = 17.77$ (York et al. 2000; Stoughton et al. 2002).  
The SDSS DR4 contains $\sim 411,000$ galaxy spectra 
distributed on several strips covering 
$\sim 4,800$ sq. degrees on the sky.
The SDSS spectroscopic sample is magnitude limited, though there
is some incompleteness due to missing nearby, 
bright galaxies ($M_{r} < 15.5$) because
of fibre saturation, missing spectra because of fibre collisions
($7$\%; Blanton et al. 2003a), redshift failures ($<1$\%), and 
missing galaxies ($\sim 1$\%) close to bright Galactic stars.  These
effects may bias results for satellites at projected
distances $r_{p} < 100$\,kpc from their hosts.  
For this reason, we perform our analysis on two subsamples.
First, we include all satellites with $r_{p} > 20$\,kpc to maximize 
the sample size while avoiding HII-region or bulge-structure contamination.
We repeat the analysis on a subsample of 
satellites with $r_{p} > 100$\,kpc, and check for consistency.
For the purposes of the present work we select two main 
samples from DR4 and split these into several subsets as we 
now specify.

Sample~1 is built using the criteria from the first sample of 
Brainerd (2005):  
(1) Hosts are at least 2.5 times (1 mag) brighter than any other galaxy
within a projected distance $r_{p} < 700$\,kpc and a line-of-sight 
velocity difference $\vert dv \vert < 1000$\,km\,s$^{-1}$;  
(2) Satellites are at least 6.25 times (2 mag) fainter than their hosts
and are found within $r_{p} < 500$\,kpc, and 
$\vert dv \vert < 500$\,km\,s$^{-1}$.

Sample~2 is constructed using the criteria of Sample~2 from 
Prada et al. (2003), but with a reduced depth of $30000$\,km\,s$^{-1}$:  
(1) Hosts are at least 2 mags brighter than any other galaxy
within $r_{p}<714$\,kpc and $\vert dv \vert < 1000$\,km\,s$^{-1}$; 
(2) Satellites are at least 2 mags fainter than hosts, are found within 
$r_{p}<500$\,kpc, and have $\vert dv \vert < 500$\,km\,s$^{-1}$ with respect 
to their hosts.  A summary of the sample selection criteria is given 
in Table~\ref{table0}.

The objects in each sample are subject to the additional 
criteria that 1) the ellipticity $e > 0.2$ 
(to ensure that the position angle is well defined), 
and 2) the primary absolute magnitude is constrained to
a bin which we set by inspecting the distribution of primary 
$r$ magnitudes. This distribution is consistent with a Gaussian with
mean $-21.99$ and standard deviation $0.85$ for Sample~1 and 
mean $-21.83$ with standard deviation $0.97$ for Sample~2.
We set the bin to $-23 < M_{r} < -21$ for both samples. 

SDSS DR4 affords larger samples than previous studies.  
Sample~1 (prior to magnitude restrictions) contains 
$4522$ satellites around $2703$ primaries, roughly 
$30$\% more than Sales \& Lambas (2004) or Brainerd (2005).  
However, we apply a magnitude selection that reduces the 
final size of Sample~1 to $3667$ satellites which is only 
slightly ($\sim 13$\%) larger than the aforementioned studies.
Though the magnitude constraint lowers our statistics,
by limiting the primary absolute magnitudes we aim to
select a small range of primary masses so that the systems
are comparable in terms of their typical dynamical states, 
masses, and formation histories.

\begin{figure}
\begin{center}
    \leavevmode
    \epsfxsize=8.0cm
    \epsfbox{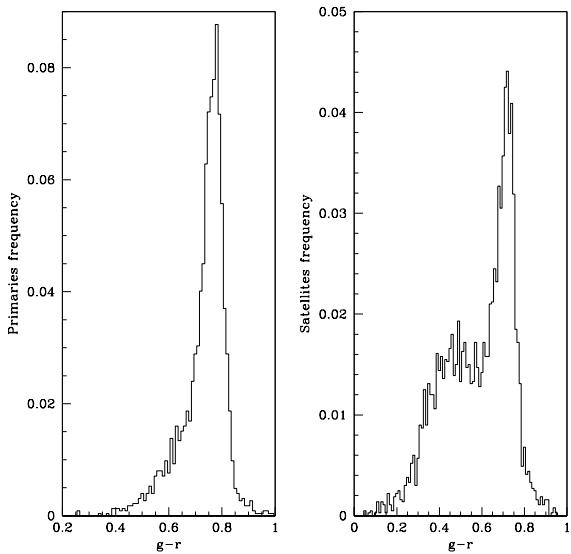}
\end{center}
\caption{The $g-r$ color distribution of the primaries and 
the satellites in Sample~1.}
\label{g-rS1}
\end{figure}

\begin{figure}
\begin{center}
    \leavevmode
    \epsfxsize=8.0cm
    \epsfbox{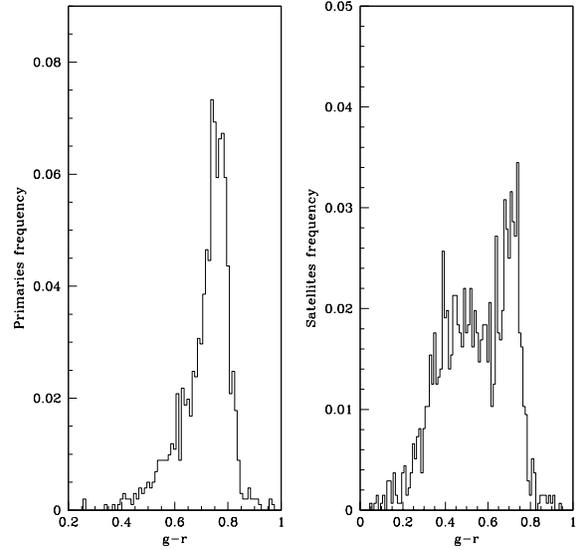}
\end{center}
\caption{Same as Fig.~\ref{g-rS1} but for Sample~2}
\label{g-rS2}
\end{figure}

We split each primary sample into subsets of Red or Blue 
satellites about either Red or Blue primaries 
(summarized in Table~\ref{table1} and Table~\ref{table3}).  
The $g - r$ color distributions of the primaries and satellites are 
shown in Fig.~\ref{g-rS1} for Sample~1 and in Fig.~\ref{g-rS2} for Sample~2.
The bimodality of the distribution is evident 
(see Strateva et al. 2001, Baldry et al. 2004), 
with the red peak dominating both the primary and satellite distributions.
The local minima between the peaks of the color distributions occur near
$g-r = 0.7$ for the primaries and $g-r = 0.6$ for the satellites,
and we use these boundaries to construct Blue and Red subsamples
of each population.

We assign each satellite an angular position $\theta$, defined as 
the angle between the primary major axis and the line joining the 
centre of the primary and the satellite.  Exploiting the assumed symmetry,
we reduce the angle $\theta$ to the first quadrant so that
$0^{\circ} \le \theta \le 90^{\circ}$.  
We take the Position Angles of the primaries from the SDSS data as the
25-mag-isophote major axis, and compute $\theta$ using this Position Angle 
and the coordinates of the objects.  
We explicitly confirmed the position angles and $\theta$ values for 
a number of random images of satellite-primary pairs.  

\begin{table}
\begin{center}
\caption{Selection criteria for Sample~1 and Sample~2. Primaries $\Delta M$
is the minimum difference in $r$ magnitude with the nearest neighbor,
Satellites $\Delta M$ is the maximum difference in $r$ magnitude with the host;
Primaries $\vert \Delta r_{p} \vert$ is the minimum projected distance
between primaries, Satellites $\vert \Delta r_{p} \vert$ is the maximum projected
distance from the host; Primaries $\vert \Delta v \vert$ is the minimum recessional
velocity difference between hosts, Satellites $\vert \Delta v \vert$ is the maximum
velocity difference with the host.}
\begin{tabular}{ccc}
Criteria & Sample~1 & Sample~2 \\
\hline
Primaries $\Delta M$ & $1.0$ & $2.0$ \\
Primaries $\vert \Delta r_{p} \vert$ & $< 700$\,kpc & $< 714$\,kpc \\
Primaries $\vert \Delta v \vert$ & $< 1000$\,km\,s$^{-1}$ & $< 1000$\,km\,s$^{-1}$ \\
\hline
Satellites $\Delta M$ & $2.0$ & $2.0$\\
Satellites $\vert \Delta r_{p} \vert$ & $< 500$\,kpc & $< 500$\,kpc \\
Satellites $\vert \Delta v \vert$ & $< 500$\,km\,s$^{-1}$ & $< 500$\,km\,s$^{-1}$ \\
\hline
\label{table0}
\end{tabular}
\end{center}
\vspace{-0.25in}
\end{table}
\begin{table}
\begin{center}
\caption{Distribution of the objects in Sample~1 (top) and Sample~2 (bottom).}
\begin{tabular}{ccc}
Color of primaries & N. of primaries & N. of satellites \\
\hline
Blue & 523 & 675 \\
Red & 1722 & 2992 \\
Tot & 2245 & 3667 \\
\hline
Blue & 312 & 378 \\
Red & 700 & 985 \\
Tot & 1012 & 1363\\
\label{table1}
\end{tabular}
\end{center}
\vspace{-0.25in}
\end{table}

\vspace{-0.15in}

\section{Results and conclusions}

\begin{figure}
\begin{center}
    \leavevmode
    \epsfxsize=8.0cm
    \epsfbox{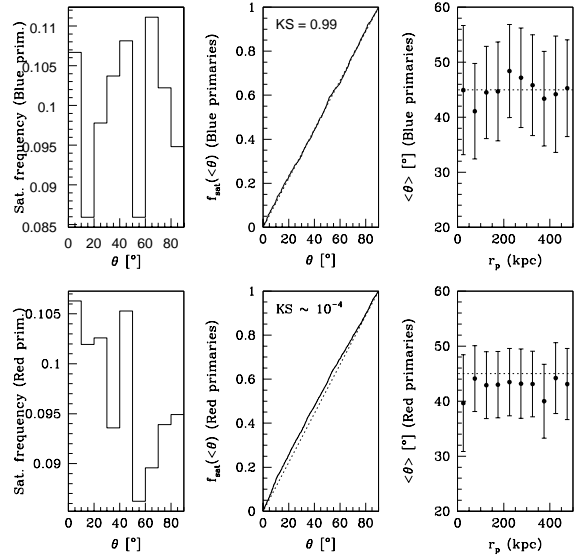}
\end{center}
\caption{Distribution of the satellites with $r_{p} > 20$\,kpc. The left
panels show the probability distribution of angular position for the
satellites of Sample~1 that reside around primaries with $-23 < M_r < -21$.  
Center panels show the cumulative angular distributions and the KS
probabilities for drawing the observed
distribution from an isotropic distribution (our null hypothesis).
Right panels show the mean angular position $\langle \theta \rangle$ 
as a function of $r_{p}$, where the errorbars indicate
the error on the mean.  The top row of panels shows
distributions for satellites about Blue primaries 
($g-r < 0.7$) and the bottom row of panels shows 
satellites around Red primaries ($g-r > 0.7$).}
\label{S1_23_21}
\end{figure}

\begin{figure}
\begin{center}
    \leavevmode
    \epsfxsize=8.0cm
    \epsfbox{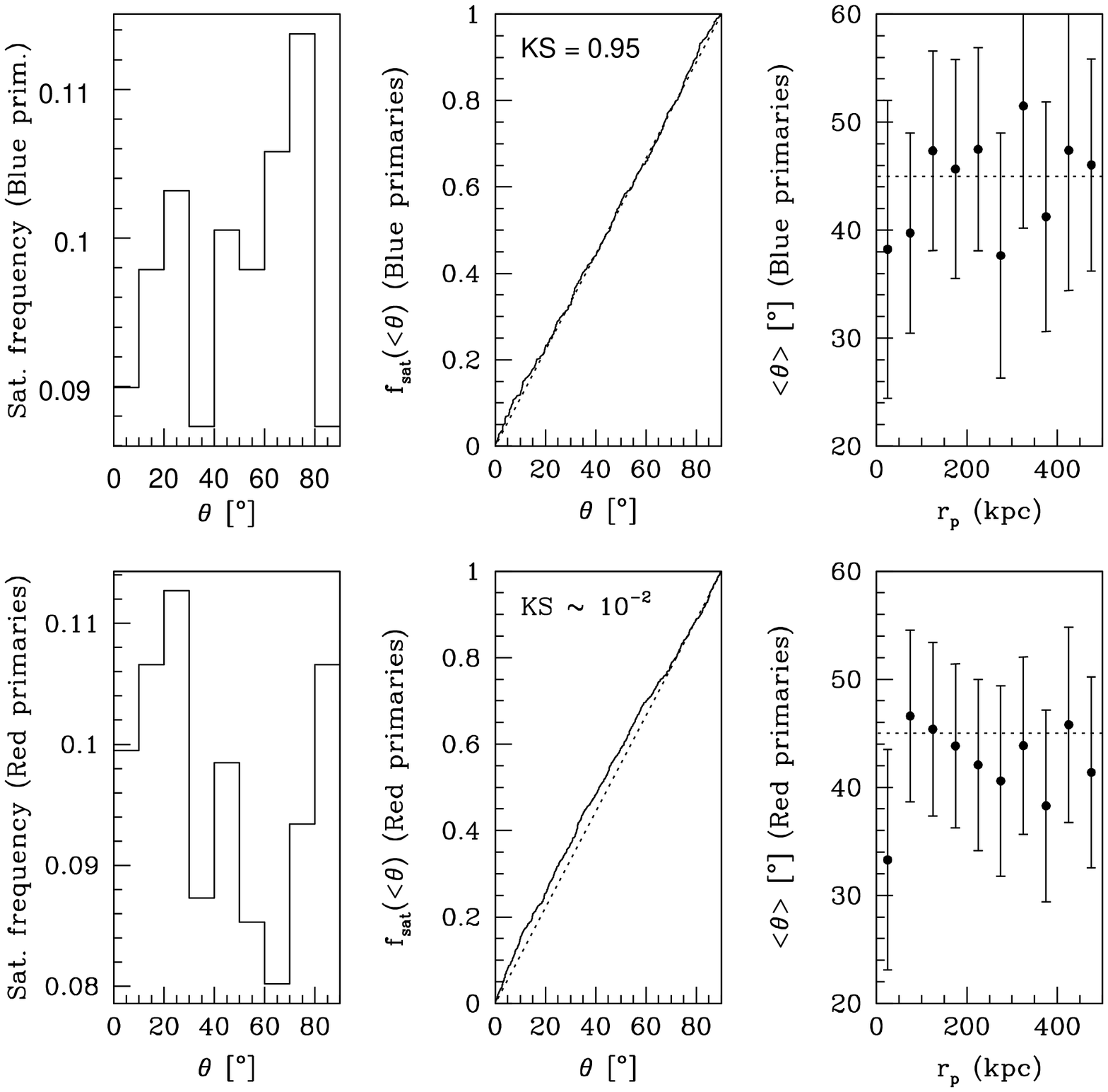}
\end{center}
\caption{Same as Fig.~\ref{S1_23_21} but for Sample~2.}
\label{S2_23_21}
\end{figure}

The angular distributions of satellites in Samples 1 and 2 
are shown in Fig.~\ref{S1_23_21} and Fig.~\ref{S2_23_21} respectively.  
The satellites of red primaries show a preferential alignment 
with the major axes of the primaries.  
This result is highly significant.  
The Kolmogorov-Smirnov (KS) probability 
of drawing the observed population of 
satellites about red primaries from an 
isotropic distribution (our null hypothesis) 
is $P_{KS} \simeq 10^{-4}$ for Sample~1.  
The satellites of blue primaries have a distribution that is 
consistent with isotropy. As a consistency test, we calculated the
average of the cosines of $\theta$.  If the angles were randomly
distributed, this average would tend to $2/\pi$.  Proximity
to this average is then translated into a probability for 
drawing the sample from an isotropic distribution.  
A summary of the results of the KS and Cosine tests
is shown in Table~\ref{table2}.
\begin{table}
\begin{center}
\caption{Results of the Kolmogorov-Smirnov and Cosine tests.
Probabilities (either KS or Cosine) are for the satellites angular distribution
to be drawn from an isotropic distribution.
The Subsets are indicated with S1 or S2 for Sample~1 or Sample~2, ``blue'' or ``red''
refer to the primary color, while ``RedBlue'' indicates the blue satellites
of red primaries, and ``RedRed'' indicates the red satellites of red
primaries. The second column shows the low limit of the projected
distance $r_{p}$ for the Subset.}
\begin{tabular}{cccc}
Subset & $r_{p}$ min. (kpc) & KS prob. & Cosine prob. \\
\hline
S1blue & 20  & $0.992$ & $0.998$ \\
S1blue & 100 & $0.979$ & $0.678$ \\
S1blue & 300 & $0.885$ & $0.979$ \\
S2blue & 20  & $0.954$ & $0.866$ \\
S2blue & 100 & $0.931$ & $0.551$ \\
S2blue & 300 & $0.613$ & $0.473$ \\
\hline
S1red & 20   &  $2.3 \times 10^{-4}$ & $2.4 \times 10^{-4}$ \\
S1red & 100  &  $7.6 \times 10^{-4}$ & $3.9 \times 10^{-4}$ \\
S1red & 300  &  $4.9 \times 10^{-3}$ & $6.9 \times 10^{-3}$ \\
S2red & 20   &  $1.0 \times 10^{-2}$ & $5.9 \times 10^{-2}$ \\
S2red & 100  &  $1.6 \times 10^{-2}$ & $3.3 \times 10^{-2}$ \\
S2red & 300  &  $0.1$                & $6.1 \times 10^{-2}$ \\
\hline
S1RedBlue & 100 & $0.302$ & $0.479$ \\
S1RedRed & 100  & $1.4 \times 10^{-4}$ & $2.0 \times 10^{-5}$ \\
\label{table2}
\end{tabular}
\end{center}
\vspace{-0.25in}
\end{table}

The satellites of the blue galaxies are consistent with isotropy 
in all cases.  To determine whether or not the sample size of 
satellites about blue primaries is large enough 
to detect a stastically-significant anisotropy 
{\it assuming} the anisotropy to be at the same level as that of the 
red sample, we drew ten random subsamples of $675$ satellites from the 
$2992$ satellites of red primaries in Sample~1.  KS tests performed on the 
random subsamples returned KS probabilities of order tens of percent 
relative to isotropic.  This is indicative that the size of 
the blue sample is insufficient to detect anisotropy assuming that 
the underlying distribution exhibits anisotropy at the same level 
as that of the red primary sample.

The right panels of Fig.~\ref{S1_23_21} and Fig.~\ref{S2_23_21} show the
mean angle $\langle \theta \rangle$, as a function of projected distance.
These figures indicate that there
is no clear dependence of $\theta$ on $r_{p}$ for any subsample.
The innermost points at $r_{p}=25$\,kpc depart from the mean
distribution, showing greater anisotropy, but this discrepancy is
not statistically significant and may be partially caused
by incompleteness in pairs at small separations and it is possible 
that such systematics may cause the feature found by Brainerd (2005) in
the same range of $r_{p}$.
To mitigate any systematics associated with pairs at small separations,
we repreated this analysis with subsamples of satellites with
$r_{p} > 100$\,kpc and confirmed the above results.

The alignment of satellites that we 
report is broadly consistent with 
that of Brainerd (2005), Yang et al. (2006), 
and Sales-Lambas (2004, after correction by Yang et al. 2006).   
Brainerd (2005) gave no indication of a color dependence for
this alignment, while Sales \& Lambas (2004) mention that 
satellites of primaries with low star-formation rates exhibit 
the strongest anisotropy, in qualitative agreement with our
alignment for satellites of red primaries.
Brainerd (2005) found that anisotropy was more
pronounced at $r_{p} < 100$\,kpc and vanished at $r_{p} > 250$\,kpc.
We find clear evidence for anisotropy in satellites with
$100 < r_{p} < 300$\,kpc and
$300 < r_{p} < 500$\,kpc.  
The satellites of the red primaries of Sample~1 
in these ranges yield $P_{KS} \simeq 5 \times 10^{-2}$ 
and $P_{KS} \simeq 5 \times 10^{-3}$ respectively.
Sales \& Lambas (2004) report a signal only when restricting to
$\vert \Delta v \vert < 160$\,km\,s$^{-1}$, whereas we apply no 
restriction in that sense yet measure significant anisotropy.   
Lastly, Yang et al. (2006) use a sample of galaxy groups, 
while we select only isolated systems.

\begin{figure}
\begin{center}
    \leavevmode
    \epsfxsize=8.0cm
    \epsfbox{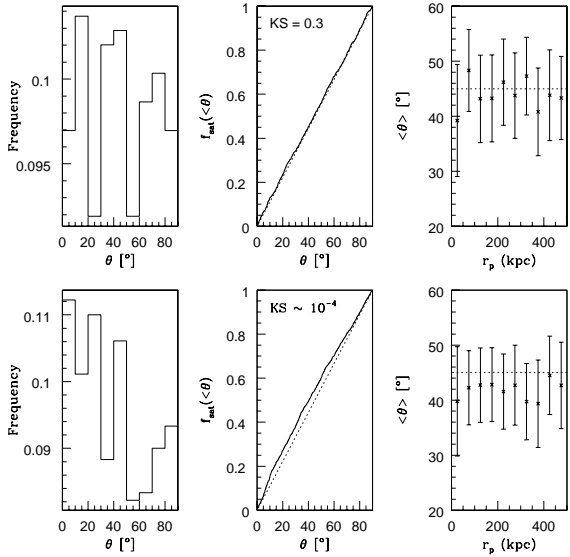}
\end{center}
\caption{
Same as Fig.~\ref{S1_23_21}, but for Blue (top row) and 
Red (bottom row) satellites of only the Red primaries of 
Sample~1.
}
\label{S1bluered}
\end{figure}

We also studied satellite angular distributions making a rough
separation into Blue and Red satellites (subsamples in Table~\ref{table3}). 
Some of the subsets contain relatively few objects, making robust, 
statistical comparisons between different populations difficult.  
Generally, red primaries have more satellites than blue primaries, 
red primaries have more red satellites than blue satellites, and 
there are few blue primaries with red satellites, as expected.  
As an example, the angular distributions of blue and red satellites about
red primaries of Sample~1 are shown in Fig.~\ref{S1bluered}.
These results extend those of Yang et al. (2006) to include isolated systems.
We find the most significant major axis alignment for
red satellites of red primaries ($P_{KS} = 1.4 \times 10^{-4}$).
We find that the satellites of blue primaries (either red or blue) 
seem to be consistent with isotropy but note that there are 
significantly fewer blue primaries.

%
\begin{table}
\begin{center}
\caption{Satellite numerical distribution in Sample~1 (top) and Sample~2 (bottom).}
\begin{tabular}{cccc}
Prim. color & Satell. color & N. of prim. & N. of satell. \\
\hline
Blue & Blue & 426 & 504 \\
Red & Blue  & 921 & 1186 \\
Blue & Red  & 141 & 168 \\
Red & Red   & 1127 & 1800 \\
\hline
Blue & Blue & 267 & 303 \\
Red & Blue  & 405 & 478 \\
Blue & Red  & 65 & 72 \\
Red & Red   & 388 & 504 \\
\label{table3}
\end{tabular}
\end{center}
\vspace{-0.25in}
\end{table}
Concerning the presence of interlopers (galaxies not bound to their
assigned primary, but classified as satellites due to projection effects)
in our samples, we could be affected by a contamination of the order of $10$\%
(as estimated by Prada et al. 2003).  However, interlopers should not 
be associated with specific primaries in any meaningful way so that they 
should only serve to dilute the anisotropy that we measure.  
In fact, Azzaro et al. (2005) demonstrated this explicitly on mock 
catalogs constructed from cosmological numerical simulations.  
Therefore, our findings should only be reinforced upon undertaking 
some (reliable) method for interloper removal.

Broadly speaking, we confirm the major axis 
alignment recently found by other authors.
In particular, we extend the results of Yang et al. (2006) to
isolated systems.  We find that angular distributions depend on both
primary and satellite color.  
Red primaries have anisotropic satellite distributions
with satellites aligned with the major axes of their hosts 
and red satellites of red primaries show the strongest 
such alignment.  Blue primaries have 
distributions consistent with isotropy.  However, we 
have shown that the sample size of blue primaries is 
insufficient to detect any anisotropy {\it assuming} an 
underlying distribution with the same level of 
anisotropy as the red primary sample.  We determine this 
by constructing random subsamples of the population of 
satellites about red primaries that had the same size as the 
sample of satellites about blue primaries and 
re-analyzing these subsamples as independent samples. 
Unlike Brainerd (2005) and Sales \& Lambas (2004),
we find a statistically-significant anisotropy for 
satellites at both large projected separations 
$r_{p} > 300$\,kpc and large relative velocities 
$\vert \Delta v \vert > 160$,km/s indicating that this 
preferential alignment is a rather general feature of 
satellites about red, or early-type primaries.
\vspace{-0.15in}

\section*{Acknowledgements}
We wish to thank Juan Betancort-Rijo for useful discussions.

SGP \& FP are supported by the
Spanish MEC under grant PNAYA 2005-07789.

ARZ is supported by the Kavli Institute for Cosmological 
Physics at The University of Chicago and by the National 
Science Foundation under grant NSF PHY 0114422.

\end{document}